\documentclass[aps,prb,twocolumn,floatfix,letterpaper]{revtex4}
\usepackage{graphicx}
\usepackage{amsmath,amssymb}

\begin{document}

\title{Phase behavior of the Lattice Restricted Primitive Model with nearest-neighbor exclusion}

\author{Alexandre Diehl}
\affiliation{Departamento de F{\'\i}sica,
Universidade Federal do Cear{\'a}, Caixa Postal 6030, CEP
60455-760, Fortaleza, CE, Brazil}
\author{Athanassios Z. Panagiotopoulos\footnote{Corresponding author: azp@princeton.edu}}
\affiliation{Department of Chemical Engineering, Princeton
University, Princeton NJ 08544, USA}

\date{\today}
\begin{abstract}

The global phase behavior of the lattice restricted primitive 
model with nearest neighbor exclusion has been studied by grand 
canonical Monte Carlo simulations.  The phase diagram is 
dominated by a fluid (or charge-disordered solid) to 
charge-ordered solid transition that terminates at the maximum 
density, $\rho^*_{max}=\sqrt2$ and reduced temperature 
$T^*\approx0.29$.  At that point, there is a first-order phase 
transition between two phases of the same density, one 
charge-ordered and the other charge-disordered. The liquid-vapor 
transition for the model is metastable, lying entirely within the 
fluid-solid phase envelope.  
\end{abstract}


\maketitle

\section{Introduction}

Recent theoretical and simulation studies of ionic systems have 
improved our understanding of their structure and thermodynamics 
\cite{Levin2004,azp2005, Fisher05, Zhou05, Ciach05, 
Hynninen05,Caillol05,Hynninen05a}. One of the most successful and 
simplest ionic models is the restricted primitive model (RPM), in 
which the ions are viewed as equisized hard spheres carrying 
positive and negative charges of the same magnitude. The RPM 
exhibits vapor-liquid phase separation and the corresponding 
critical point was confirmed to belong to the three-dimensional 
Ising universality class~\cite{Luijten02,Kim04}. 

The discretized version of the RPM, the lattice restricted 
primitive model (LRPM), has also been extensively studied by both 
simulation 
\cite{Dickman99,azp99,Almarza01,Diehl03,Diehl05,Hoang05} and 
theoretical approaches 
\cite{Dickman99,Ciach,Ciach03,Ciach04,Ciach04b,Moghaddam05,Artyomov05}. 
In the LRPM, positions of the positive and negative ions of 
diameter $\sigma$ are restricted to the sites of an underlying 
simple cubic lattice of spacing $l$; the parameter 
$\zeta=\sigma/l$ specifies how closely the lattice system 
approaches the continuum behavior. In addition to the obvious 
computational advantages~\cite{azp99}, since the interactions 
between all lattice sites are pre-computed, the lattice model 
presents some unusual characteristics generally absent from 
non-ionic fluids~\cite{azp00}. For $\zeta=1$, the most striking 
feature is an order-disorder transition~\cite{Dickman99,azp99}, 
which is not present in the continuous version of the RPM. There 
is no gas-liquid transition and the coexistence is between a 
low-density disordered phase and an antiferromagnetically ordered 
high-density phase.  The transition is continuous (N\'eel-type 
line) above and first-order below a tricritical point. However, 
for fine discretized lattices with $\zeta\geqslant 3$ the 
vapor-liquid coexistence is recovered and the critical point and 
coexistence curves converge to the values found in the continuum 
model for increasing values of $\zeta$~\cite{Kim04,azp99,azp02}.

Several investigations have been performed of the $\zeta = 1$ 
LRPM with additional short-range attractions~\cite{Ciach,Diehl03} 
or nearest-neighbor (nn) 
repulsions~\cite{Ciach03,Ciach04,Diehl05}. These models present a 
rich phase behavior as the nn strength is varied. For weak or 
vanishing nn interactions only order-disorder transitions and a 
tricritical point were found, while for increasing nn strength 
different scenarios could be possible, depending on the magnitude 
of the nn interaction. For short-range attractions both 
tricritical and gas-liquid critical points can become 
stable~\cite{Diehl03}, while for nn repulsion the continuum-space 
behavior is recovered~\cite{Diehl05}.

The present paper extends the previous study~\cite{Diehl05}, of 
the LRPM model with variable nearest neighbor repulsion to the 
limit of infinite repulsion.  The model is thus the lattice 
restricted primitive model with nearest neighbor exclusion 
(LRPM-nn).  This is equivalent to a lattice restricted primitive 
model of discretization parameter $\zeta = \sqrt2$.  Prior 
theoretical studies~\cite{Ciach03,Ciach04} with first, second and 
third nn exclusion indicate close similarity between the LRPM-nn 
phase diagram and the results for an off-lattice ionic system at 
high densities. Of particular interest to the present work are 
possible connections between order-disorder transitions and lower 
density vapor-liquid transitions and high density transitions 
between charge-ordered and charge-disordered phases
\cite{Ciach04}. The present paper is organized as follows. The 
model and computational details are given in Sec.~\ref{model}. 
Results are discussed in Sec.~\ref{results}. We close in 
Sec.~\ref{conclusions} with summary and conclusions.

\section{Model and simulation methods}
\label{model}

We consider a system of $2N$ charges, half of them carrying 
charge $+q$ and half charge $-q$, on a simple cubic lattice. 
We enforce nearest neighbor exclusion on the lattice and set the 
charge diameter as the unit of length, $\sigma = 1$. The lattice 
spacing is then $l=1/\sqrt 2$, such that the lattice discretization 
parameter~\cite{azp99} is defined as $\zeta = \sigma/l=\sqrt 2$.  The 
charges interact through the (continuous space) Coulomb potential
\begin{equation}
\label{potential} 
U_{ij}= \frac{q_i q_j}{D r_{ij}}\;,  
\end{equation}
where $D$ is the dielectric constant of the structureless solvent 
in which the charges are immersed. While some studies of the 
phase behavior of the true lattice Coulomb potential are 
available~\cite{Kobelev03}, most prior studies of lattice RPM 
models have been done using the potential of equation (\ref{potential}).  

Reduced quantities are defined as follows:
\begin{equation}
\label{reduced} 
T^{\ast}= \frac{k_B T}{E_0}, \quad  
\rho^{\ast}=\frac{2N\sigma^3}{V}, \quad U^{\ast}=\frac{U} {E_0} \;,
\end{equation}
where $\sigma$ is the ion diameter (taken as the unit of length), 
$V$ is the volume of the system, $U$ the energy per ion pair and 
$E_0=q^2/D\sigma$ is the magnitude of Coulomb energy between two 
ions at close contact. The reduced chemical potential, 
$\mu^{\ast}$, is defined so that at the limit of high 
temperatures and low densities, $\mu^{\ast}\to 2T^{\ast}\ln N\sigma^3/V$, 
where the factor 2 comes from the presence of two 
ions per minimal neutral ``molecule'' inserted or deleted in the 
simulations.  With this choice of reduced units, the maximum 
density of the system at close packing is $\rho^{\ast}_{\rm max}=\sqrt{2}$.

The reduced box length, $L^{\ast}=L/\sigma$ is a non-integer 
quantity, as the lattice spacing is $l=1/\sqrt2$; it is more 
convenient to use the lattice spacing as the reducing length in 
this case, so we define 
\begin{equation}
\label{L_dag} L^{\dag}= \frac{L}{l}=\sqrt2\frac{L} {\sigma}\;. 
\end{equation}

The LRPM-nn model studied in the present work is equivalent to 
the limiting case of infinite repulsive coupling to 
nearest neighbor sites, $J \rightarrow \infty$, in the model 
studied in Ref.~\cite{Diehl05}.  However, Ref.~\cite{Diehl05} used an 
effective charge diameter of $\sigma =\sqrt{2}$, so that reduced 
densities and temperatures are higher in the present study by 
factors of $2\sqrt{2}$ and $\sqrt{2}$, respectively. The unit 
conventions used here facilitate comparisons with previous 
studies of the LRPM model~\cite{azp99} and the continuous 
RPM~\cite{Bresme00,Vega03}.

Electrostatic interactions were computed using Ewald summation 
with conducting boundary conditions at infinity, 518 
Fourier-space wave vectors and real-space damping parameter 
$\kappa =5$. Interactions were pre-computed at the beginning of 
the simulation runs for all possible pairs of lattice sites and 
stored in an array for computational efficiency.

We used grand canonical Monte Carlo (GCMC) simulations in cubic 
boxes subject to full periodic boundary conditions.  Two types of 
moves were utilized, namely pair additions and removals and 
swapping of oppositely charged ions to enhance sampling of 
order-disorder transitions. To enhance acceptance of the 
insertion and removal steps we used distance-biased sampling, 
following Ref.~\cite{orkoulas94}. Particle swaps constituted up 
to 60 \% of attempted moves, depending on temperature and density.

Multihistogram reweighting~\cite{azp02,Ferrenberg88,Frenkel} 
techniques were used to analyze the simulation data. For the 
critical region we used mixed-field finite size scaling analysis 
proposed by Bruce and Wilding~\cite{Wilding92}, which accounts 
for the lack of symmetry between coexisting phases in fluids. We 
did not attempt to incorporate corrections for pressure mixing in 
the scaling fields, as any such effects are expected to be 
small~\cite{Young04}. The Tsypin and Bl\"{o}te~\cite{Tsypin00} 
limiting distribution for the three-dimensional Ising model was 
used for obtaining the critical parameters. Typical runs 
involved $10^8$ Monte Carlo steps for equilibration and 
$10^9$ steps for production.  Such runs required approximately 10 
CPU hours on 3 GHz Pentium 4 processors.  Longer runs were 
performed near the vapor-liquid critical point and at high 
densities at which the acceptance ratio for insertions and 
removals was lower.  Statistical uncertainties were obtained from 
multiple independent runs with different pseudorandom sequences.  
The random number generator ``ran2'' of Ref.~\cite{recipes} was 
used for the calculations.

\begin{figure}
\includegraphics[width=8cm]{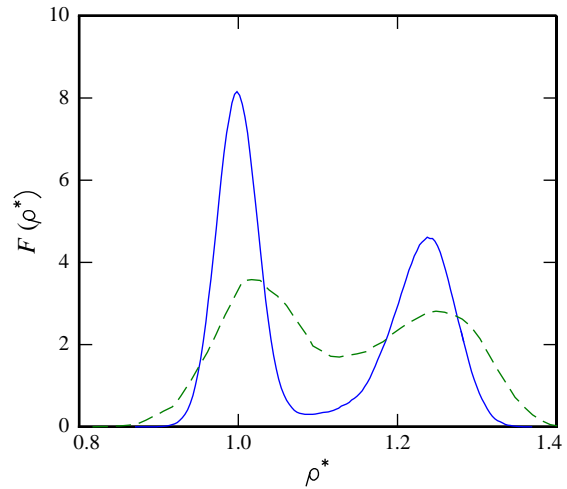}
\caption{(color online) Normalized density distribution, 
$F(\rho^{\ast})$, at $T^{\ast}=0.15$ in simulation boxes of size 
$L^{\dag}=12$ (continuous line) and $L^{\dag}=8$ (dashed line). } 
\label{fig1}
\end{figure}
For the charge-disordered solid to charge-ordered solid part of 
the phase diagram (or ``fluid''-solid), we were able to observe 
direct transitions between the ordered and disordered phases, as 
shown in Fig.~\ref{fig1}.  This established the relative free 
energies of the two phases and eliminated the need for 
thermodynamic integration to a reference state of known free 
energy.  As seen in Fig.~\ref{fig1}, for smaller boxes the 
probability of intermediate densities is greatly enhanced and the 
system readily passes between the two phases.  For simulation 
boxes much greater than $L^{\dag}=12$, sampling is restricted to 
the phase from which the simulation is started; it is not 
possible to establish a reversible path from the 
charge-disordered solid (or fluid) to the charge-ordered solid 
phases using grand canonical simulations without umbrella 
sampling.

\section{Results and discussion}
\label{results} 

\begin{figure}
\includegraphics[width=8cm]{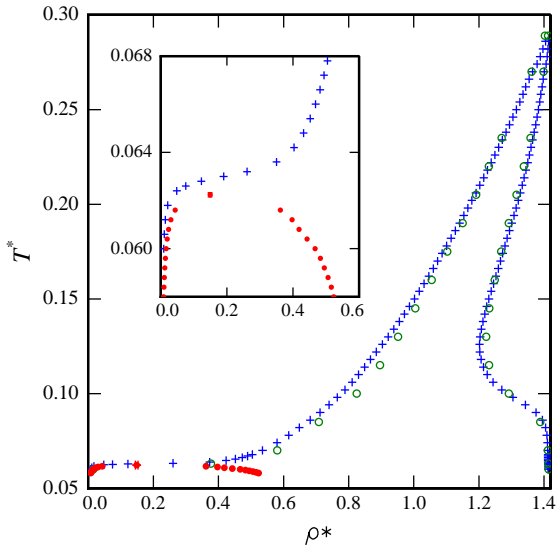}
\caption{(color online) Phase behavior of the LRPM-nn model.  The 
``fluid''-solid transition is indicated by crosses 
($L^{\dag}=12$) and open circles ($L^{\dag}=8$).  Filled circles 
mark the vapor-liquid equilibrium phase boundaries obtained in a 
box of size $L^{\dag}=24$.  The inset expands the region around 
the metastable vapor-liquid critical point. Statistical 
uncertainties are smaller than symbol size.} \label{fig2}
\end{figure}

Fig.~\ref{fig2} shows the overall phase behavior for the LPRM-nn 
model.  The ``fluid''-solid transition dominates the phase 
diagram.  It should be pointed out that the designation ``fluid'' 
is not appropriate at high densities.  There is no first-order 
transition between fluid and solid phases in the $\sqrt2$ lattice 
hard sphere model to which the present model reduces at the limit 
of high temperatures.  The $\sqrt2$ lattice hard sphere model has 
a second order transition at a density of 
$\rho^*\approx0.59$~\cite{azp05,Gaunt67}, above which an 
fcc-ordered sublattice exists in the system.  The lower-density 
phase for the LPRM-nn is a disordered fluid at lower temperatures 
but becomes increasingly solid-like at higher densities.  Near 
the end-point of the transition the coexisting phases are both 
solids with face-centered-cubic overall arrangement of the 
particles, if one ignores the ion charge.  For this reason, we 
enclose the term ``fluid'' in quotation marks when referring to 
the phase at the low-density side of this transition, to 
acknowledge the fact that the phase changes continuously from a 
disordered fluid to a charge-disordered solid.

The vapor-liquid envelope is seen in the inset to Fig.~\ref{fig2} 
to be metastable and wholly within the ``fluid''-solid boundary. 
This is surprising for a system with long-range (Coulombic) 
interactions because the lack of a stable liquid phase is usually 
associated with \emph{short-range} attractions~\cite{Wolde97}.  
This behavior can be rationalized by considering that the 
presence of the underlying lattice greatly stabilizes the solid, 
thus shifting the fluid-solid transition to higher temperatures.  
By contrast, for systems with very short range interactions, it 
is the destabilization of the liquid phase that leads to the 
disappearance of the vapor-liquid transition.  The phase diagram 
for the LRPM-nn model is similar to that of the LRPM with 
$\zeta=2$, obtained with large statistical uncertainties in 
Ref.~\cite{azp99} and studied theoretically in \cite{Ciach04b}.

There is some system size dependence observed for the 
``fluid''-solid part of the phase diagram, especially at lower 
temperatures, as seen from the difference of the apparent phase 
boundaries.  The smallest of the system sizes shown in 
Fig.~\ref{fig2} ($L^{\dag}=8$) can accommodate only 256 ions at 
full packing; it was included only for comparison purposes.  The 
larger system ($L^{\dag}=12$) can accommodate 864 ions at close 
packing, but is still small enough to allow for efficient direct 
sampling of the order-disorder transition, as seen in 
Fig.~\ref{fig1}.

\begin{figure}
\includegraphics[width=8cm]{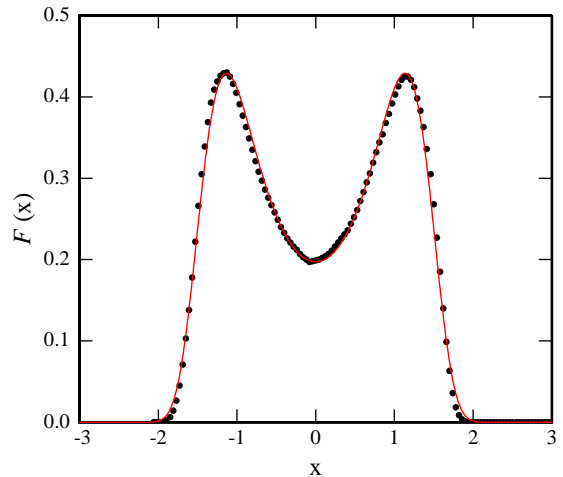}
\caption{(color online) . Order parameter distribution, $F(x)$, 
for a system of $L^{\dag}=24$ at conditions corresponding to the 
critical point listed in Table I.  Points are simulation data and 
the line represents the analytical approximation of 
Ref.~\cite{Tsypin00} for the three-dimensional Ising universality 
class limiting distribution. Simulation statistical uncertainties 
are comparable to symbol size.} \label{fig3}
\end{figure}
The liquid-vapor critical point and phase coexistence envelope 
were obtained for system sizes ranging from $L^{\dag}=14$ to 
$L^{\dag}=24$.  Because of the lower densities relative to the 
``fluid''-solid transition, larger systems were required.  
Results for the critical parameters are shown in Table I.  The 
apparent (system-size dependent) critical temperature, 
$T_{c}^{\ast}$, chemical potential, $\mu_{c}^{\ast}$ and field 
mixing parameter $s^{\ast}_{\rm mix}$ were obtained by minimizing 
deviations between the universal order parameter 
distribution~\cite{Tsypin00} and the observed distributions.  A 
typical optimized order parameter distribution is shown in 
Fig.~\ref{fig3} , in which the abscissa is the mixed-field order 
parameter, $x=N(U^{\ast}-s^{\ast}_{\rm mix})$.   The estimates 
for the critical temperature do not vary significantly with 
system size, while the critical density first increases and then 
becomes slightly lower with system size.

\begin{table} \label{tab1}
\caption {Vapor-liquid critical parameters.  Numbers in 
parentheses indicate statistical uncertainties in units of the 
last figure shown.} \vspace{0.3cm}
\begin{tabular} {|cc|cccc|} \hline
& $L^{\dag}$ &  $~T_{c}^{\ast}$  & $\mu_{c}^{\ast}$   & $s^{\ast}_{\rm mix}$  & $\rho_{c}^{\ast}$ \\
\hline
& 15         &~~0.0623(1)       & -1.5015(1)         & -0.643(1)  & 0.132(1) \\
& 18         &~~0.0624(1)       & -1.5021(1)         & -0.621(1)  & 0.155(2) \\
& 22         &~~0.0622(1)       & -1.5016(2)         & -0.630(5)  & 0.150(4) \\
& 24         &~~0.0622(1)       & -1.5041(1)         & -0.629(3)  & 0.148(3)  \\
\hline
\end{tabular}
\end{table}

These critical parameters can be compared with those for the 
related model of Ref.~\cite{Diehl05} with partial exclusion of 
nearest neighbor sites, $J=0.3$, after taking into account the 
different reducing parameters. The $J=0.3$ model has critical 
parameters (for $L^{\dag}=15$) $~T_{c}^{\ast}=0.0612(1)$, 
$\rho_{c}^{\ast}=0.14(2)$, very similar to the LPRM-nn present 
model which corresponds to $J \rightarrow \infty$.  However, the 
high-density (solid) phases of the two models are completely 
different, as discussed in the paragraphs that follow.  

The critical temperature for the present model is slightly higher 
and the critical density significantly higher than the continuum 
RPM, for which $~T_{c}^{\ast}=0.04933(5)$, 
$\rho_{c}^{\ast}=0.075(1)$ \cite{Kim04}.  Vapor-liquid critical 
parameters for models with $\zeta=$ 3, 4 and 5 have been obtained 
in \cite{azp99}; the critical temperatures and densities were 
found to be higher in coarser lattices, a trend consistent with 
the results of the present study.  

At relatively high temperatures, Coulombic interactions become 
less important than excluded volume in determining the equation 
of state for the LPRM-nn model.  At the limit of very high 
temperatures, we have already mentioned that the model is 
equivalent to a $\sqrt2$ lattice hard sphere model.  At higher 
densities an fcc-ordered solid appears and at close packing every 
sphere occupies an ordered position.  In LRPM-nn, the 
high-temperature solid is substitutionally disordered with 
respect to charge type. The continuous RPM 
model~\cite{Bresme00,Vega03,Vega96} has a transition from a 
charge-disordered to a charge-ordered solid phase at temperatures 
near $T^{\ast}\approx 0.29$.  The fully occupied LRPM 
charge-ordered to charge-disordered phase transition has been 
investigated previously \cite{Almarza01} and found to have a 
first order transition at the same temperature.  The phase 
behavior of Fig.~\ref{fig1} has the same transition at a similar 
temperature range.  

\begin{figure}
\vspace{1cm}
\includegraphics[width=5cm]{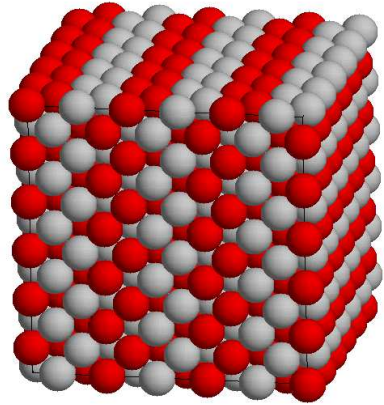}
\caption{(color online) Structure of the solid, $L^{\dag}=12$. 
Different shades (colors) represent oppositely charged ions.} 
\label{fig4} 
\end{figure}
The structure of the ion-ordered solid is shown in 
Fig.~\ref{fig4}.   The structure is \emph{P4/mmm} (tetragonal), 
identical to the ``fcc''-ordered structure observed at high 
densities for the continuum RPM \cite{Vega96,Bresme00,Vega03}.  
This structure has not yet been observed experimentally in 
systems of oppositely charged colloids \cite{Hynninen06}.

\begin{figure}
\includegraphics[width=8cm]{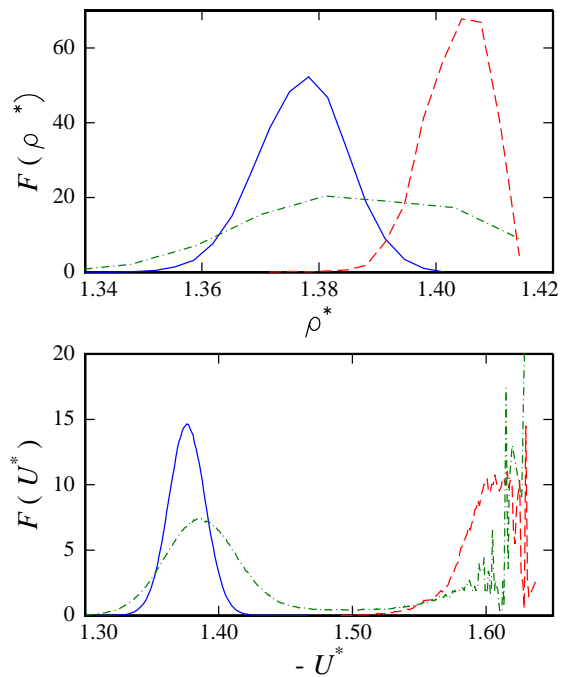}
\caption{(color online) Density (top) and energy (bottom) 
distributions, $F(\rho^{\ast})$ and $F(U^{\ast})$ at 
$T^{\ast}=0.28$, $\mu^{\ast}=0.8$.  Solid line: $L^{\dag}=12$, 
started from disordered solid; dashed line: $L^{\dag}=12$, 
started from ordered solid; dash-dotted line: $L^{\dag}=8$.} 
\label{fig5}
\end{figure}
In Ref.~\cite{Abascal03}, the character of the transition between 
ordered and disordered solid phases for the continuous RPM was 
investigated using constant-pressure Monte Carlo simulations and 
found to be ``weakly first order.''  For the LRPM-nn model, the 
densities of the coexisting phases (ordered and disordered 
solids) are seen in Fig.~\ref{fig2} to converge at the maximum 
density $\rho^*_{max}=\sqrt2$.  Normalized probability 
distributions for the densities and energies are shown in 
Fig.~\ref{fig5} at $T^{\ast}=0.28$ for two system sizes, 
$L^{\dag}=12$ and $L^{\dag}=8$.  The apparent ``noise'' at low 
energies (to the right of the bottom part of the graph) is due to 
the finite number of states with one, two etc vacancies in the 
lattice model.  For the smaller box, densities and energies 
corresponding to both phases are sampled in a single run.  There 
is no hint of two peaks in the density distribution, but the 
energy distribution shows a clear separation of states into 
higher energy (less negative, disordered) and lower energies 
(ordered).  For the larger box size, the simulations get trapped 
in the phase from which they are started; even though the density 
distributions of the two runs at identical chemical potentials 
overlap at $\rho^{\ast}\approx 1.39$, there is no overlap in the 
energy distributions.

\begin{figure}
\includegraphics[width=8cm]{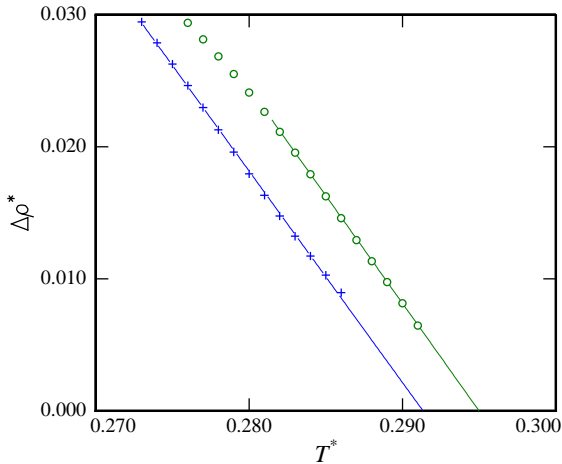}
\caption{(color online) Density difference between 
charged-ordered and charged-disordered phases for $L^{\dag}=12$ 
(crosses) and $L^{\dag}=8$ (open circles).  Statistical 
uncertainties are comparable to symbol size.  Lines are 
linear-least-squares fits to the points over the range 
indicated.} \label{fig6}
\end{figure}

We have computed the density difference between the 
charged-ordered and charged-disordered phases near the transition 
end-point by using energy (rather than density) to identify the 
phases.  In other words, referring to Fig.~\ref{fig5}, we 
collected the density under the charge-ordered (more negative 
energy) and charge-disordered (less negative energy) phases.  The 
results allow us to extend the coexistence envelope to higher 
temperatures for which the density differences are small.  The 
results for the density difference, $\Delta\rho^*$, as a function 
of temperature are shown in Fig.~\ref{fig6}.  The 
linear-least-squares fits to the points are indicated as lines in 
Fig.~\ref{fig6}. These give in turn the temperature at which the 
first order transition occurs with \emph{no density 
discontinuity}, as $T^{\ast}=0.295(2)$ for $L^{\dag}=8$ and 
$T^{\ast}=0.291(3)$ for $L^{\dag}=12$.  At that limit, both 
coexisting phases are at the closed-packed density, 
$\rho^*=\rho^*_{max}=\sqrt2$.  A fluid-solid phase transition 
with no density discontinuity has been established for the 
Gaussian core model \cite{Prestipino05}.  Experimentally, several 
metallic elements, in particular Ce, Cs, Ba and Eu, have melting 
lines of zero slope in the pressure-temperature plane 
\cite{Kennedy62,Jayaraman63,Jayaraman65}, also indicating a 
fluid-solid transition with no density discontinuity.

It is of interest to compare our results to the field-theoretic 
study of Ciach and Stell \cite{Ciach04} for the LPRM-nn and to 
simulations of Abascal \emph{et al.} \cite{Abascal03} and Bresme 
\emph{et al.} \cite{Bresme00} for the continuous RPM at high 
densities.  The main difference between our results and these 
prior studies is that we find that the charge-ordered and 
charge-disordered solids have a first-order transition with zero 
density difference at the closed-packed density.  We speculate 
that such a transition may also be found in the continuous RPM 
near the close-packed density, as our findings seem to be 
generally consistent with those of Ref.~\cite{Abascal03}.

\section{Conclusions}
\label{conclusions}

We have studied the phase diagram of the lattice restricted 
primitive model with nearest neighbor exclusion (LPRM-nn), using 
grand canonical Monte Carlo simulation and histogram reweighting 
techniques.  The global phase diagram is dominated by the 
``fluid''-solid transition, which starts with a large density gap 
between a dilute gas phase and the solid at low temperatures.  
The transition ends at $T^{\ast}=0.291(3)$ as a first order 
transition between charge-ordered and charge-disordered phases of 
the same density, $\rho^*=\sqrt2$.  First-order transitions 
between phases of the same density in one-component systems have 
been observed for several metallic elements and for the Gaussian 
core model.

The liquid-vapor phase transition for the model was determined to 
be metastable, lying entirely within the ``fluid''-solid phase 
envelope.  The metastable critical point for the transition was 
obtained as $~T_{c}^{\ast}=0.0622(1)$, $\rho_c^* = 0.148(3)$, 
values higher than for the continuum RPM but consistent with 
previously determined trends for discretized lattice models.

While the broad outline of the phase diagram is consistent with 
theoretical predictions \cite{Ciach04}, our results differ from 
these predictions in some important aspects.  In particular, the 
liquid / fcc-disordered transition is not present in our system.  
Our results are in near-quantitative agreement with calculations 
of ordered-disordered fcc phase transitions for the continuum RPM 
\cite{Abascal03}.  However, we find that the charge-ordered and 
charge-disordered phases maintain a first-order transition even 
though there is no density difference between the coexisting 
phases.

\section*{Acknowledgments}
AD acknowledges financial support of the Brazilian agency CNPq - 
Conselho Nacional de Desenvolvimento Cient\'\i fico e 
Tecnol\'ogico. AZP acknowledges funding by the Department of 
Energy, Office of Basic Energy Sciences (through Grant No. 
DE-FG02-01ER15121) and ACS-PRF (Grant 38165 - AC9).  We are 
grateful to Dr. Frank Stillinger for pointing out the existence 
of pure component systems that have first-order transitions with 
no density discontinuity. We also would like to thank Dr. Alina 
Ciach, Dr. Carlos Vega, and Dr. Antti-Pekka Hynninen for helpful 
comments and discussions.

\end{document}